\documentclass[twoside,twocolumn,9pt]{article}
\usepackage[super,sort&compress,comma]{natbib} 
\usepackage[version=3]{mhchem}
\usepackage[left=1.5cm, right=1.5cm, top=1.785cm, bottom=2.0cm]{geometry}
\usepackage{balance}
\usepackage{mathptmx}
\usepackage{sectsty}
\usepackage{graphicx} 
\usepackage{lastpage}
\usepackage[format=plain,justification=justified,singlelinecheck=false,font={stretch=1.125,small,sf},labelfont=bf,labelsep=space]{caption}
\usepackage{float}
\usepackage{fancyhdr}
\usepackage{fnpos}
\usepackage[english]{babel}
\addto{\captionsenglish}{%
  
}
\usepackage{array}
\usepackage{droidsans}
\usepackage{charter}
\usepackage[T1]{fontenc}
\usepackage[usenames,dvipsnames]{xcolor}
\usepackage{setspace}
\usepackage[compact]{titlesec}
\usepackage{hyperref}
\usepackage{siunitx}

\usepackage{epstopdf}

\definecolor{cream}{RGB}{222,217,201}

\begin{document}

\pagestyle{fancy}
\thispagestyle{plain}
\fancypagestyle{plain}{
\renewcommand{\headrulewidth}{0pt}
}

\makeFNbottom
\makeatletter
\renewcommand\LARGE{\@setfontsize\LARGE{15pt}{17}}
\renewcommand\Large{\@setfontsize\Large{12pt}{14}}
\renewcommand\large{\@setfontsize\large{10pt}{12}}
\renewcommand\footnotesize{\@setfontsize\footnotesize{7pt}{10}}
\makeatother

\renewcommand{\thefootnote}{\fnsymbol{footnote}}
\renewcommand\footnoterule{\vspace*{1pt}%
\color{cream}\hrule width 3.5in height 0.4pt \color{black}\vspace*{5pt}} 
\setcounter{secnumdepth}{5}

\makeatletter 
\renewcommand\@biblabel[1]{#1}            
\renewcommand\@makefntext[1]%
{\noindent\makebox[0pt][r]{\@thefnmark\,}#1}
\makeatother 
\renewcommand{\figurename}{\small{Fig.}~}
\sectionfont{\sffamily\Large}
\subsectionfont{\normalsize}
\subsubsectionfont{\bf}
\setstretch{1.125} 
\setlength{\skip\footins}{0.8cm}
\setlength{\footnotesep}{0.25cm}
\setlength{\jot}{10pt}
\titlespacing*{\section}{0pt}{4pt}{4pt}
\titlespacing*{\subsection}{0pt}{15pt}{1pt}

\fancyfoot{}
\fancyfoot[LO,RE]{\vspace{-7.1pt}\includegraphics[height=9pt]{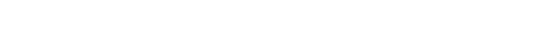}}
\fancyfoot[CO]{\vspace{-7.1pt}\hspace{13.2cm}\includegraphics{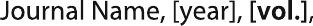}}
\fancyfoot[CE]{\vspace{-7.2pt}\hspace{-14.2cm}\includegraphics{RF}}
\fancyfoot[RO]{\footnotesize{\sffamily{1--\pageref{LastPage} ~\textbar  \hspace{2pt}\thepage}}}
\fancyfoot[LE]{\footnotesize{\sffamily{\thepage~\textbar\hspace{3.45cm} 1--\pageref{LastPage}}}}
\fancyhead{}
\renewcommand{\headrulewidth}{0pt} 
\renewcommand{\footrulewidth}{0pt}
\setlength{\arrayrulewidth}{1pt}
\setlength{\columnsep}{6.5mm}
\setlength\bibsep{1pt}

\makeatletter 
\newlength{\figrulesep} 
\setlength{\figrulesep}{0.5\textfloatsep} 

\newcommand{\topfigrule}{\vspace*{-1pt}%
\noindent{\color{cream}\rule[-\figrulesep]{\columnwidth}{1.5pt}} }

\newcommand{\botfigrule}{\vspace*{-2pt}%
\noindent{\color{cream}\rule[\figrulesep]{\columnwidth}{1.5pt}} }

\newcommand{\dblfigrule}{\vspace*{-1pt}%
\noindent{\color{cream}\rule[-\figrulesep]{\textwidth}{1.5pt}} }

\makeatother

\twocolumn[
  \begin{@twocolumnfalse}
{\includegraphics[height=30pt]{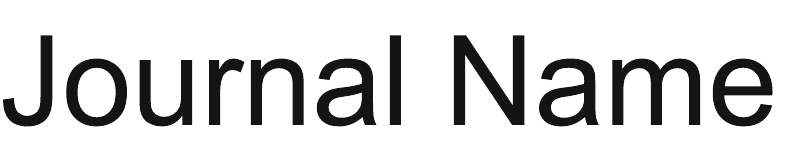}\hfill\raisebox{0pt}[0pt][0pt]{\includegraphics[height=55pt]{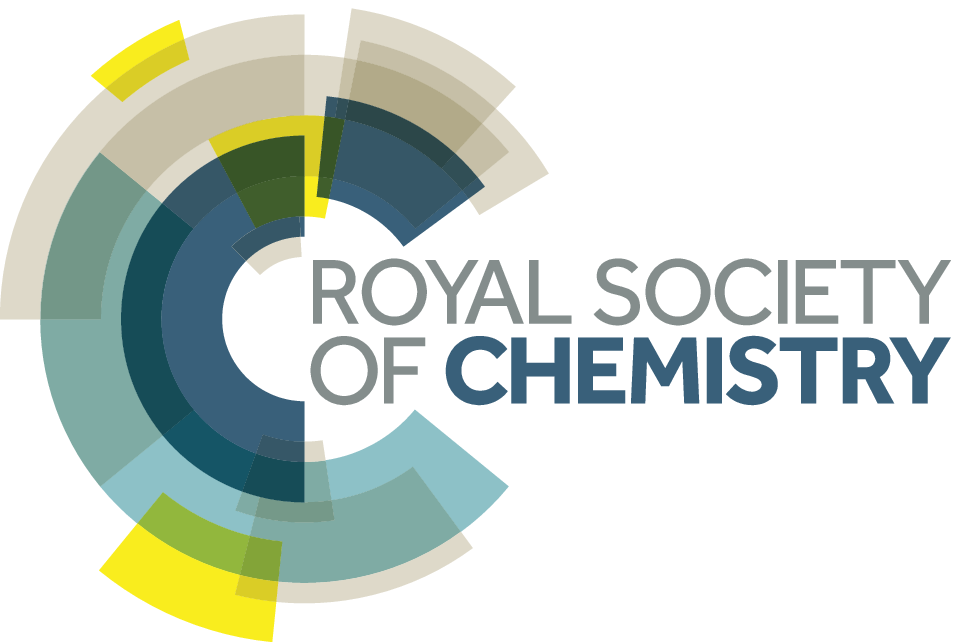}}\\[1ex]
\includegraphics[width=18.5cm]{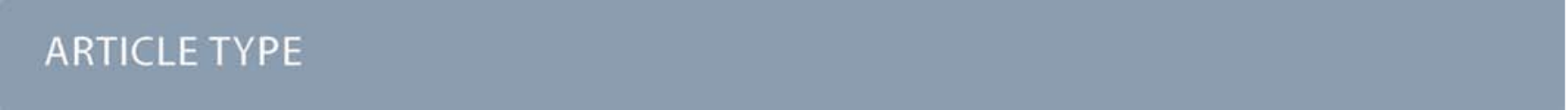}}\par
\vspace{1em}
\sffamily
\begin{tabular}{m{4.5cm} p{13.5cm} }

\includegraphics{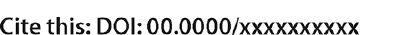} & \noindent\LARGE{\textbf{Quantum Rate as a Spectroscopic Methodology for Measuring the Electronic Structure of Quantum Dots$^\dag$}} \\
\vspace{0.3cm} & \vspace{0.3cm} \\

 & \noindent\large{Edgar Fabian Pinzón,\textit{$^{a}$} Laís Cristine Lopes,\textit{$^{a}$} André Fonseca,\textit{$^{b}$}} Marco Antonio Schiavon,\textit{$^{b}$} and Paulo Roberto Bueno,$^{\ast}$\textit{$^{a}$} \\

\includegraphics{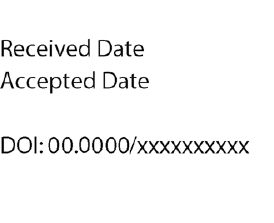} & \noindent\normalsize{The electronic structure of nanoscale moieties (such as molecules and quantum dots) governs the properties and performance of the bottom-up fabricated devices based on their assemblies. Accordingly, simple and faster experimental methods to resolve the electronic density of states of these nanoscale materials (of which quantum dots are a particular example) are of great importance for the development of man-made nanoscale interfaces and nanoelectronics. In the present work, we propose the quantum rate spectroscopy methodology (and introduce the fundamental physical basis of this technique that was previously demonstrated for measuring the electronic structure of two-dimensional compounds such as graphene~\cite{Lopes-2024}) as a tool for resolving the electronic structure of zero-dimensional (quantum dot) structures at room temperature and environmental pressure conditions. This method is simpler than the traditional methods based on scanning tunneling microscopy. This spectroscopic approach based on the quantum rate theory was demonstrated for CdTe quantum dots and was used to measure a spectrum that provides discrete energy levels that are consistent with those obtained by tunneling microscopy measurements.} 

\end{tabular}

 \end{@twocolumnfalse} \vspace{0.6cm}

  ]

\renewcommand*\rmdefault{bch}\normalfont\upshape
\rmfamily
\section*{}
\vspace{-1cm}

\makeatletter\def\Hy@Warning#1{}\makeatother

\footnotetext{\textit{$^{a}$Sao Paulo State University, Rua Francisco Degni, 55 - Araraquara, Sao Paulo, Brazil}}
\footnotetext{\textit{$^{b}$The Federal University of São João del-Rei, Praça Frei Orlando, 170 - São João del-Rey, Minas Gerais, Brazil }}

\footnotetext{\dag~Electronic Supplementary Information (ESI) available: [details of any supplementary information available should be included here]. See DOI: 00.0000/00000000.}

\footnotetext{\ddag~Additional footnotes to the title and authors can be included \textit{e.g.}\ `Present address:' or `These authors contributed equally to this work' as above using the symbols: \ddag, \textsection, and \P. Please place the appropriate symbol next to the author's name and include a \texttt{\textbackslash footnotetext} entry in the the correct place in the list.}



\section{Introduction}

\begin{figure*}[t!]
\centering
\includegraphics[height=8.0cm]{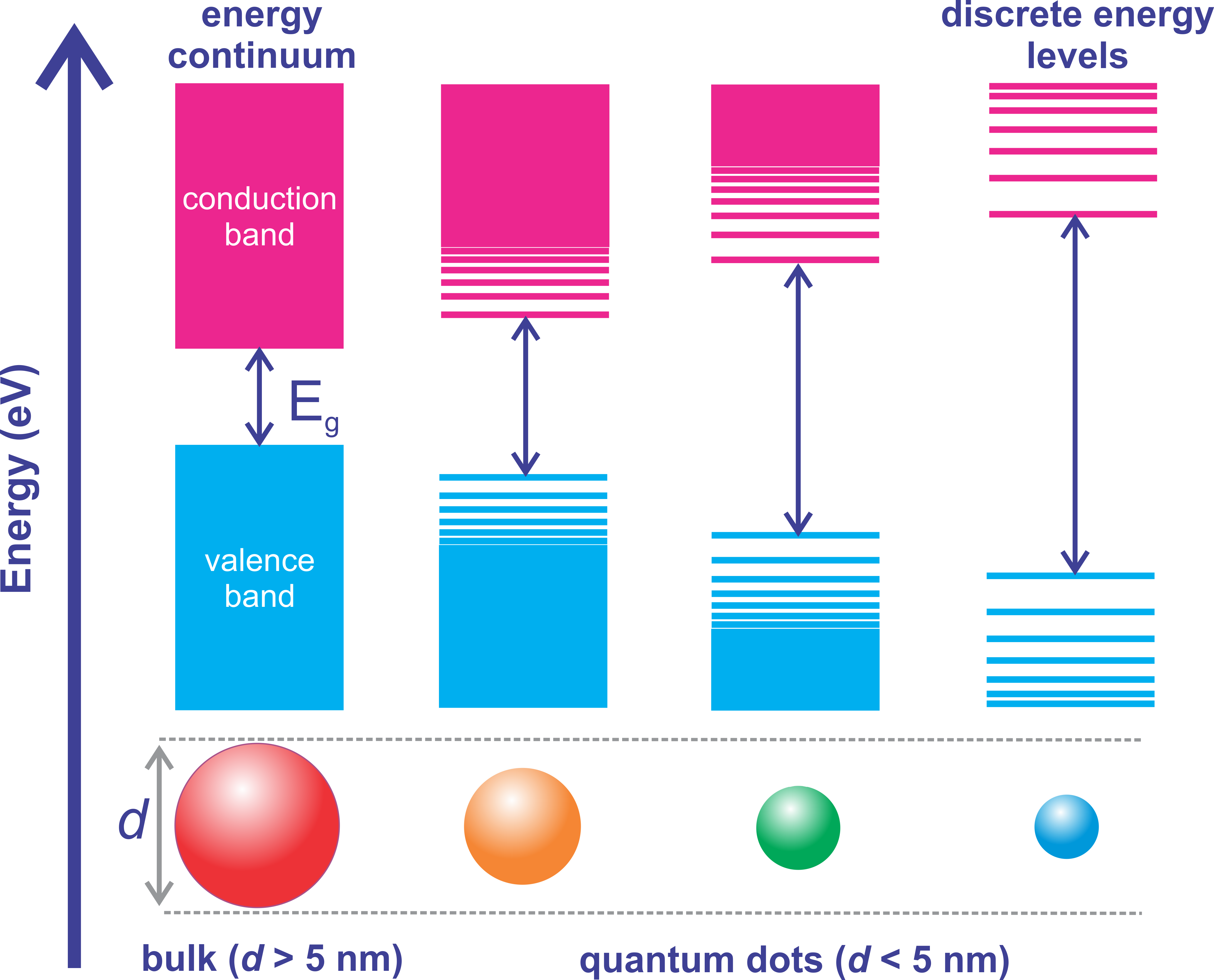}
\caption{Schematic representation of the energy levels for low-dimensional QD semiconducting materials. Note the evolution of the quantum confinement of the energy levels as a function of the dimension of the QD. For instance, the QD with dimensions greater than 5 nm (on the left) has a pattern that is more similar to a bulk semiconducting material in which the electronic structure is formed by two separated bands (termed conduction and valence bands and the separation between these bands is termed the band gap because it constitutes an energy interval in which electron occupation is prohibited) within the internal energy levels that are quite close to each other, forming a continuum of characteristic energy levels referred as the band structure. As the size of the QDs decreases below 5 nm, the separation between the energy levels in these bands increases, forming a set of discrete (rather than a continuum) energy states that at a certain size resembles the energy levels observed in molecules (scheme at right). Note also that the band gap of semiconducting materials increases with decreasing particle size due to the above-described quantum confinement effect. In this figure 5 nm size of QD was used as a reference. The precise cut-off depends on the definition of the Bohr radius of a QD. For instance, for CdTe QDs this value is calculated as $\sim$ 6.8 nm~\citep{QDs-classification-2021}.}
\label{fig:confinement}
\end{figure*}

Charge carriers dynamics in semiconducting quantum dot (QD) structures are restricted in all of the three possible Cartesian geometric directions at the nanoscale in a range of 2-10 nm. Accordingly, QDs are referred to as zero-dimensional structures due to their spatial point characteristics governed by the above-mentioned nanoscale size features. This unique feature of QDs makes them the closest man-made semiconducting inorganic constructions with electronic structure comparable to that of organic molecules, as noted in Fig~\ref{fig:confinement}. Hence,  QDs provide discrete quantized energy levels accessible at room temperature. Accordingly, QDs are special materials that offer excellent optical and electronic properties\cite{Akbari2020} that are unique compared to those of microscopic structures or bulk materials\cite{Edvinsson2018}. 

The electronic structure of molecules or nanoscale spatially-restricted atomic assemblies is often analyzed in terms of their electron density of states (DOS) that is a function that accounts for the amount of quantum-mechanical states of electrons (or electron energy levels) per interval of energy normalized by the volume of the atomic structure. Therefore, the DOS is a fingerprint of the electronic structure of atomic and molecular assemblies that is an important tool for predicting the properties of man-made nanoscale materials, particularly when the DOS can be experimentally measured\cite{Houtepen2005}.

Techniques commonly applied for experimental study of the electronic properties of QDs are based on optical excitations and energy level transitions that can be easily applied to the investigations of the electronic transition properties of nanocrystals. Useful information about the influence of the size and shape on the properties of these nanoscale crystals can be deduced by optical absorbance, luminescence, and excitation spectroscopic methods\cite{Jdira2006}. However, the inherent limitations of these techniques do not allow a full investigation of the DOS; rather only some energy levels can be investigated in an indirect manner. 

For the scanning tunneling microscopy (STM) and spectroscopy (STS) methods, it is challenging to separately investigate the electron and hole energy levels, and additionally the electron beam used to obtain the information can damage the nanostructures and cause undesirable atomic strain and extrinsic defects\cite{Erni2008}. Although investigations of the local density of states (LDOS) of single nanoparticles have been performed using these methods, it is experimentally challenging to discern the variation of LDOS within and between QDs\cite{Bakkers2001}.

The use of STS to characterize CdSe QDs of different sizes is a direct approach to  determine the relationship between the energy levels and the size of QDs\cite{Jdira2006}. A combination of STS with optical spectroscopy demonstrated that the relative rates of the tip-to-dot and dot-to-substrate tunneling currents influence the electron occupancy in the QDs. Therefore, the tunneling current has been measured as a function of the tip-substrate potential bias difference controlled via the STS feedback settings. A large number of experiments for each set-point of electric current were carried out to successfully improve the reproducibility of the results. The transitions in the tunneling spectra were explained by a master equation accounting for the electron-hole occupancy of the discrete energy levels ascribed to the conduction and valence band states (or orbitals) of CdSe QDs.

Even though STM- and STS-based methodologies have made significant contributions to the on-going understanding of the electron energy levels of QDs, the wide use of these methodologies is either not practical or requires unique and expensive equipment that is not easy to operate because of the required high-vacuum and/or low-temperature conditions. Therefore, experimental approaches that are both simpler and have bench-top capability are desired to advance the characterization of the electronic density of low-dimensional structures.

Therefore, the goal of the present work is to demonstrate a proof-of-principle of a methodological approach, named quantum rate spectroscopy (QRS), as a simpler method of probing the electronic structure of nanoscale chemical assemblies of interest of material scientists, chemists, and physicists. The use of QRS was successfully applied to reveal the electronic structure of two-dimensional compounds such as graphene~\cite{Lopes-2024}. This electrochemical QRS method of measuring the electronic structure of graphene was demonstrated to be in good agreement with the electronic structure measured angle-resolved photo-emission spectroscopy and that theoretically computed by density-functional theory~\cite{Lopes-2024}. QRS is an electric spectroscopy method that is fundamentally based on the quantum rate theoretical principles\citep{Bueno2020, Bueno-book-2018, Bueno-2023-3}. For instance, besides the application of the principles to characterize single-layer graphene\cite{Lopes2021, Lopes-2024}, it has also been possible to access the DOS of mixed metal-oxide films\cite{Pinzon2021} and organic assemblies\cite{Feliciano2020}, such as molecular redox-active films\cite{Bueno-2023-3}. Furthermore, quantum rate theory has been useful for the study of quantum electrochemical principles~\citep{Alarcon2021, Bueno-2023-3}, for example demonstrating that charge transfer resistance is essentially a specific setting of the quantum conductance principle~\cite{Sanchez-QR-Rct}. This has provided the basis for the design of improved electrochemical interfaces for electron transport~\cite{Sanchez-QR-efficience-2022} with multiple applications.

In this work, the QRS methodology is demonstrated as a useful electric spectroscopic tool for evaluating the electronic structure of CdTe QD assemblies. Two different nanocrystal sizes (2.25 and 3.27 nm) are studied and the obtained experimental spectra are mathematically fit to a quantum mechanical statistical equation following the quantum rate principles, hence demonstrating that the measured electronic spectra are in agreement with the quantum rate theory predictions. Furthermore, it is demonstrated that the energy levels obtained in these CdTe QD spectra are in agreement with those obtained by other studies, for example by using the STM method.

\section{Experimental Section}
\subsection{Chemical Reagents and Solutions}
CdCl$_2$$\cdot$H$_2$O (99$\%$) was purchased from Vetec. 3-mercapropropionic acid (MPA; 98$\%$), NaOH (99$\%$), Na$_2$TeO$_3$ (99$\%$), NaBH$_4$ (99$\%$), and Rhodamine 6G (99$\%$) were obtained from Sigma Aldrich. Acetone (98$\%$) was obtained from Synth. All chemicals were used as received, without further purification. Milli-Q water was used for all experiments. Phosphate buffer (PB) solution (12 mM PB, pH 7.4) was prepared using 0.2 M KNO$_3$, 10 mM Na$_2$HPO$_4$$\cdot$12 H$_2$O, and 2 mM KH$_2$PO$_4$.

\subsection{Synthesis of CdTe Quantum Dots} 
Cadmium telluride (CdTe) QDs were synthesized via the One-Pot colloidal method. Initially, CdCl$_2$.H$_2$O (0.40 mmol) and surface ligand (3 mercaptopropionic acid MPA, 0.80 mmol) were solubilized in Milli-Q ultrapure water (80 mL) in a three-neck flask and the pH of this solution was adjusted to 10.00 with 1.0 mol.L$^{-1}$ sodium hydroxide. Then, Na$_2$TeO$_3$ (0.04 mmol) and NaBH$_4$ (0.1 mmol) were added and the system was heated to 100 $^{\circ}$C under reflux to keep the precursors at constant concentrations. When the temperature reached 100 $^{\circ}$C, aliquots were taken out at different time intervals to measure the sizes of the quantum dots. After the synthesis, the QD dispersion was purified by the precipitation method, using acetone as a non-solvent to destabilize the colloidal suspension. Then, the precipitate was collected by centrifugation at 9000 rpm for 5 min and resuspended in Milli-Q ultra-pure water.

\subsection{Characterization of CdTe Quantum Dots}
Absorption spectra were acquired in the 200–700 nm region using a spectrophotometer (Shimadzu, UV-2450/2550, Japan) with a spectral resolution of 1.0 nm and a slit width of 1.0 nm. UV-Vis absorption spectra were acquired using a spectrophotometer (Shimadzu, UV-2450/2550, Japan), with a spectral resolution of 1.0 nm, and a slit width of 1.0 nm. Photoluminescence (PL) spectra were obtained using a spectrofluorometer (Shimadzu, RF-5301 PC, Japan) equipped with a 150 W Xenon lamp. The spectral resolution utilized was 1.0 nm, the excitation and emission slits were both set to 3.0 nm and the excitation wavelength to 355 nm. All spectroscopy studies were performed using 10.00 mm quartz cuvettes (Shimadzu) at room temperature. Powder X-ray diffraction (XRD) was performed with a diffractometer (XRD-600, Shimadzu, Japan) using CuK$\alpha$ radiation with a 0.3 mm divergence slit. The diffraction angle (2$\theta$) was varied from 10 to 60 degrees at a rate of 1$^{\circ}$ min$^{-1}$ and a step of 0.02$^{\circ}$. The results obtained from the characterization of the CdTe QDs are presented in detail in the SI.

\subsection{Preparation of the Assemblies of Quantum Dots}
Gold electrodes (Au electrodes) were fabricated by radio-frequency sputtering. Briefly, titanium and gold were sequentially deposited over silicon oxide wafers (0.8 x 0.5 cm$^2$) obtaining gold electrodes with an average geometric area of approximately 0.04 cm$^{2}$. These Au electrodes were initially cleaned via sonication periods of 12 min following the use of a sequence of solvents, namely isopropyl alcohol, acetone, isopropyl alcohol, and ultrapure water. Then, these surfaces were cleaned by 30 min under ultraviolet/ozone. The pre-treated Au electrodes were incubated in a solution containing 50 mmol L$^{-1}$ of L-cysteine during 16 h at room temperature to form a L-cysteine monolayer over the surface of the Au. After this chemical modification of the surface, the electrode was rinsed with ultrapure water. The L-cysteine monolayer-modified Au electrodes were incubated in an aqueous solution containing 0.2 mmol L$^{-1}$ of CdTe QDs, 100 mg mL$^{-1}$ EDC and 100 mg mL$^{-1}$ NHS during 3 h in the dark. The EDC and NHS are added in the solution to activate the carboxyl groups (COOH) of the QDs surface, allowing their conjugation with the amine groups (NH$_2$) of the L-cysteine (contained in the surface of the Au electrode) through amide bonding. Finally, the CdTe-QD-modified Au electrodes were rinsed with ultrapure water.

\subsection{Electrochemical Measurements}
Electrochemical impedance spectroscopy (EIS) was performed using a portable potentiostat (PalmSens4) equipped with a frequency response analyzer (FRA) in a three-electrode setup. The CdTe QDs assembled over Au electrodes served as the working electrode, Ag|AgCl (3M KCl) as the reference electrode and a platinum mesh as the counter electrode. All electrochemical measurements were carried out in a phosphate buffer solution (pH 7.4). EIS spectra were obtained at a fixed bias potential (which in the case of this study was 0.14 V \textit{versus} Ag|AgCl (3M KCl)), corresponding to the value of the open-circuit potential, OCP) using an amplitude potential perturbation of 10 mV and frequencies ranging from 1 MHz to 0.05 Hz. 

The obtained raw EIS data were used to calculate real $C'$ and imaginary $C''$ capacitive components by applying the following relationship $Z^{*}(\omega) = 1/j\omega C^{*}(\omega)$, where $Z^{*}(\omega)$ and $C^{*}(\omega)$ are the impedance and capacitance complex functions, respectively. $\omega$ is the perturbing angular frequency used in the measurements. 

The complex capacitance $C^{*}(\omega)$ function has a particular meaning within the quantum rate theory~\cite{Lopes2021}

\begin{equation}
 \label{eq:Cq-complex}
	C_q^*(\omega) = \frac{C_q}{1 + j\omega \tau} \sim C_q \left( 1 - j\omega \tau \right) + O(\omega^2),
\end{equation}

\noindent corresponding to a relaxation function that can be described in terms of the  equivalent quantum circuit elements. For instance, this relaxation function is expressed in terms of a series combination of $R_q$ and $C_q$, providing a characteristic frequency $\omega_c$ with a time $\tau = 2\pi R_q C_q$ (or $\omega_c = 1/R_q C_q$) that directly provides information regarding the quantum rate $\nu = 1/\tau$ as defined further in Eq.~\ref{eq:nu}.

\begin{figure*}[t!]
\centering
\includegraphics[height=4.2cm]{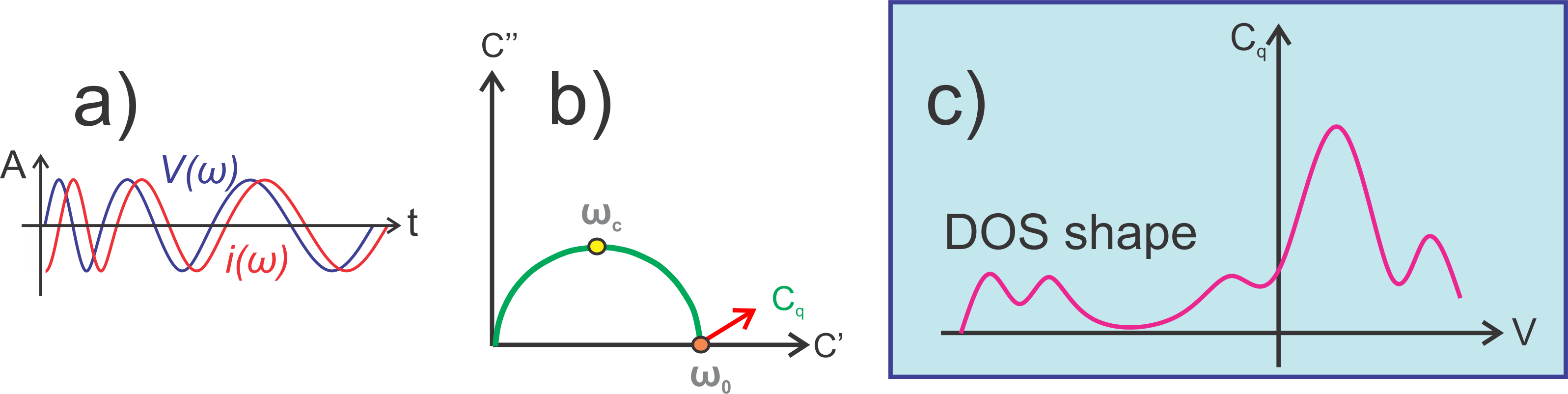}
\captionof{figure}{Schematic representation of the measurements of the DOS of QDs based on the quantum rate principles as can be obtained from EIS measurements. (a) EIS measurement consists of a sinusoidal potential or energy perturbation of the QDs that provide an electric current response to this perturbation from which a complex impedance function $ \left[ Z^{*}(\omega) \right]$ is obtained. From $Z^{*}(\omega)$, a complex capacitive function $ \left[ C^{*}(\omega) \right] $ is obtained and plotted as shown in a capacitive (b) Nyquist diagram, enabling the measurement of the equilibrium charge occupancy. This equilibrium charge occupancy is obtained from the capacitive Nyquist diagram as a particular low-frequency value from which $C_q$ is taken, corresponding to the value of the diameter of the semi-circle in the diagram. Once the equilibrium angular frequency $\omega_0$ is obtained as the frequency at which $C_q$ is obtained, a potential scan of $C_q$ at $\omega_0$ provides the (c) DOS pattern of a QD ensemble assembled at the interface. Note that in (b) $\omega_c = 1/R_qC_q$ refers to the characteristic angular frequency, which is directly related to the quantum rate as described in Eq.~\ref{eq:nu}.}
\label{fig:EIS-methodology}
\end{figure*}

Therefore, noting that the complex quantum capacitive relaxation function $C_q^*(\omega)$ can be expressed as $C_q^*(\omega) = C' + j C''$ and that $C' = \varphi Z''$ and $C'' = \varphi Z'$, where $\varphi = 1/\omega|Z|^2$, it can be observed that $C'$ and $C''$ can be obtained from EIS measurements. From the  EIS spectra, real $Z'$ and imaginary $Z''$ impedance components can be readily measured. $|Z|$ is the impedance modulus that can be calculated from $Z'$ and $Z''$ components. The frequency associated with the maximum value of the real capacitive term, i.e., corresponding to the diameter of the Nyquist capacitive semicircle (see Fig.~\ref{fig:EIS-methodology}\textit{b}), which is theoretically obtained for $\omega \rightarrow 0$ in Eq.~\ref{eq:Cq-complex}, is experimentally obtained as a finite value and is fixed as a constant charge-occupancy equilibrium state. By fixing this $\omega_0$ limit frequency value and performing a potential scan (which in the present study ranged from -1.8 to 1.8 V \textit{versus} Ag|AgCl (3M KCl), the  $C_q(\omega_0)$ function is obtained. This $C_q(\omega_0)$ plotted as function of potential energy at the fixed $\omega_0$ of the electrode is used to obtain the shape of the DOS by noting that $C_q \propto$ DOS (see theory discussed below). The QRS method is illustrated in Fig.~\ref{fig:EIS-methodology}. Note that all potentials used in the present work are reported \textit{versus} Ag|AgCl (3M KCl) reference.

\section{Results and Discussion}

Before describing how the electronic DOS of an ensemble of semiconducting QDs is measured using QRS, it is useful to introduce the key concepts of the QRS method, which is based on the quantum rate theory, as mentioned above.

\subsection{Fundamentals of Quantum Rate Theory}\label{sec:QRS-fundamentals}

We start by assuming that an ensemble of QDs with an average size lower than 5 nm (with an electronic structure such as shown in Figure~\ref{fig:DOS}) can be assembled over a conducting electrode (see Fig.~\ref{fig:electrode-QDs}). Therefore, the electronic states (including valence and conduction band states) are accessible by perturbing electrons and holes by a sinusoidal wave potential or energy $E$ perturbation imposed from the electrode in which $E = h \nu$ is the quantum mechanical principle that specifies the relationship between energy $E$ and frequency $\nu$ (in hertz) that can be applied to investigate the DOS of a QD ensemble assembled on the electrode. Note that following the usual notation, $h$ is the Planck's constant.  

One of the key assumptions of the quantum rate theory~\citep{Bueno-2023-1, Bueno-2023-2, Bueno-2023-3} is that it allows us to model electron particles and information exchanged between two different quantum mechanical states and hence to better understand the $E = h \nu$ dynamics phenomenon at super or extremely low energies perturbation, i.e. down to 30 Hz, corresponding to energies lower than 124 feV, as it is the case of redox reactions~\citep{Bueno2020, Bueno-2023-3} and of the long-range electron transport in respiration chains~\cite{Bueno-2024}. For inorganic quantum dots, similarly to the case of graphene~\citep{Bueno-2022, Lopes-2024}, as will be demonstrated in the present work, this lower energy perturbation phenomenon obeys quantum mechanical rules that cannot be correctly modeled using Schr\"ordinger's non-relativistic quantum-wave methods but can be using relativistic quantum electrodynamics within Planck-Einstein $E = h \nu$ relationship.

Note that the quantum electrodynamics related to the quantum rate theory is different of the common sense for relativistic electrodynamics in which it is guessed that a particle must to be close to the velocity of a photon. It must be clarified that this note the case. Quantum rate theory implies quantum electrodynamics and Planck-Einstein relationship because the principles of the theory complies with the Dirac equation~\citep{Dirac-1928}, which was an equation formulated exactly with the purpose to comply with Planck-Einstein relationship wherein the Schrödinger equation fails, though it is a particular setting of the Dirac equation whenever spin dynamics is disregarded. In general, the Dirac equation implies a wave description of the electron in which the massless character is implicit. The meaning of this is that electrons can, in certain circumstances, behave like waves by presenting a null rest mass $m_0$. This massless Fermionic characteristic implies relativistic behaviour simply because it follows a particular setting of the relativistic relationship $E^2 = p^2 c_*^2 + m_0^2 c_*^4$, where for $m_0 = 0$ it leads to $E = pc_*$, not only obeying the Planck-Einstein relationship but also complying with the Dirac quantum mechanical equation. 

In other words, the quantum rate theory defines a fundamental rate $\nu$ principle based on the ratio between the reciprocal of the von Klitzing constant $R_k = h/e^2$ and the quantum (or chemical) capacitance, such as~\citep{Bueno-2023-3}

\begin{equation}
 \label{eq:nu}
	\nu = \frac{e^2}{hC_q} = \frac{E}{h},
\end{equation}

\noindent where by noting that $E = e^2/C_q$ is an energy intrinsically associated with the electronic structure, it straightforwardly leads to the Planck-Einstein relationship, i.e.

\begin{equation}
 \label{eq:Planck-Einstein}
	E = h\nu = \hbar \textbf{c}_* \cdot \textbf{k},
\end{equation}

\noindent which follows a linear relationship dispersion between the energy $E$ and wave-vector\footnote{Note that \textbf{k} in bold refers to the wave-vector and its magnitude will be referred to as |\textbf{k}| to avoid notation issues with the meaning of $k$ that, in previous works~\citep{Bueno-2023-3,Sanchez-QR-Rct}, was referred to as the electron transfer rate constant.} $\textbf{k}$, where $\textbf{c}_*$ is the Fermi velocity and $\hbar$ is the Planck constant $h$ divided by $2\pi$. Hence, the quantum rate principle within Eq.~\ref{eq:nu} predicts relativistic dynamics that comply with Dirac~\citep{Dirac-1928} instead of with the non-relativistic Schr\"ordinger equation. Note that the linear relationship between \textbf{k} and $E$ is intrinsically relativistic which is demonstrable by invoking De Broglie relationship, which states that the momentum $\textbf{p} = \hbar \textbf{k}$ is directly related to the $\textbf{k}$; hence Eq.~\ref{eq:Planck-Einstein} turns into $E = \textbf{p} \cdot \textbf{c}_*$, demonstrating its intrinsic relativistic character.

It has been experimentally demonstrated~\citep{Alarcon2021, Bueno-2023-3, Sanchez-QR-Rct} that $E = e^2/C_q$ is a degenerated state of energy for experiments conducted in an electrolyte environmental medium, such as that effectively $E$ is $E = g_s g_e (e^2/hC_q)$, where $g_s$ is the electron spin degeneracy and $g_e$ is the energy degeneracy state associated with the electric-field screening effect of the electrolyte over the quantum states contained in molecules\cite{Bueno-2023-3}, graphene\citep{Bueno-2022, Lopes-2024}, as well as quantum dots (as will be demonstrated here).

There are two equivalent interpretations of $g_e$ degeneracy. The first is based on the existence of resonant electric currents, i.e., there are time-dependent (displacement) electric currents within the junction. This type of resonant electric current is owing to the existence of two charge carriers (electrons and holes) that promote a net current denoted here as $i_0$. The origin of this displacement electric current is owing to the role played by the electrolyte that allows the superimposition of the electrostatic $C_e$ and quantum capacitive $C_q$ modes of charging the molecular junction states. In this situation, the elementary charge $e$ is subjected to an equivalent electric potential, i.e., $e/C_e \sim e/C_q$, conducting locally to $C_e \sim C_q$. 

Hence, the equivalent $C_\mu$ capacitance of the junction is $1/C_\mu = 2/C_q$, with an energy degeneracy of $e^2/C_\mu = 2e^2/C_q$, where 2 is accounted as $g_e$ in the formulation of $\nu = g_e G_0/C_q$ concept, as stated in Eq.~\ref{eq:G/C-degeneracy}~\cite{Bueno-2023-3}. This energy degeneracy is equivalent to the previous electrical current degeneracy consideration of the origin of $g_e$ because $1/C_\mu = 2/C_q$ implies $C_q = 2C_{\mu}$. Noting that the capacitance of the interface is a result of the equivalent contribution of $C_e$ and $C_q$, there is an electric current degeneracy for charging the quantum capacitive states of the interface that is proportional to $C_\mu$ with an electric current of $i_0 = C_{\mu}s = (1/2)(C_q)s$, which is equivalent to $2 i_0 = C_q s$, where $s = dV/dt$ is the time-dependent potential perturbation (scan rate) imposed to the interface to investigate the energy $E = e^2/C_q$ level, electronically coupled to the electrode states. Note that it is owing to the equivalence of $e^2/C_e$ and $e^2/C_q$ energy states that there is an energy degeneracy that inherently permits two electric currents contributing to the net (electrons and holes) current $i_0$ of the interface, which is thus said to be a ambipolar electric current.

The next section will introduce how the above concepts permit us to establish a low-energy electrodynamics that allows us to spectroscopically measure (in an electrolyte and room temperature environment) the electronic structure of ensembles attached to an electrode.

\subsection{Quantum Rate Spectroscopic: Method and Principles}\label{sec:QRS-method}

Spectroscopic methods based on $\nu = E/h$ phenomenon within Eq.~\ref{eq:nu} are well-known. The `novelty' brought by the quantum rate theory is noting that a spectroscopic method can be applied if the quantum capacitance $C_q$ is experimentally accessible at super or extremely low-frequencies, where the electronic communication with the quantum states of the QDs can be established using a perturbation imposed by the electrode that serves as a probe or a frequency source for a spectroscopic method based on this principle. This can be achieved in a range of super or extremely low frequencies $\nu$, and communication with the quantum energy states of the QDs is possible owing to the electronic coupling of the QD states with those of the electrode within an electrolyte environment. The electronic coupling $\kappa$ can be of different natures and intensities, according to the chemical way of coupling the moieties -- molecules or QDs, for instance -- to the electrodes, but it will allows a rate frequency such as $\nu = \kappa E/h$ to measurable and hence the QRS method will apply, as will be demonstrated further here.

Note that $E = h \nu$ has a particular setting in this spectroscopic methodology, where the energy $E$ of the accessible states is related to the quantum capacitive $C_q$~\cite{Bueno-David-2016} states of the QDs such as $E \propto e^2/C_q$ and $\nu$ is the quantum rate, i.e., the rate (the inverse of the characteristic time constant) at which the $E$ states of the QDs are communicating with those of the electrode. 

Equivalently to the case of graphene~\citep{Lopes-2024}, the information about the QD electronic density of states is accessible through the measurement of the quantum capacitance $C_q$~\cite{Bueno2020, Bueno-David-2016} in an electrochemical experimental setting without requiring high-vacuum or low-temperature conditions such as those required in the case of STM/STS method, for instance. Note that $C_q$ directly reports on the density of states DOS $= C_q/e^2 = (dn/dE)$ of the molecular compounds assemblies over the electrode. Nonetheless, the presence of an electrolyte is vital because it allows the suppression of the Coulomb effect~\cite{Pinzon-2022} and permits the direct measurement of $C_q$ of the QDs, as will be demonstrated below. This QRS methodology allows us to define the rate at which the states of the QDs communicate with the electrode owing to an electric-field perturbation of the QD states through the electrode. Theoretically, this quantum rate can be equivalently defined as the ratio between the quantum of conductance $G$ and the electrochemical capacitance $C_\mu$ as, given by~\cite{Feliciano2020, Bueno-2023-3, Bueno-book-2018}

\begin{equation}
\label{eq:nu-2}
\nu = \frac{G}{C_\mu}= g_s \frac{e^2}{h}\sum_{n=1}^{N} T_n(\mu) \left(\frac{1}{C_e} + \frac{1}{C_q}\right),
\end{equation}

\noindent where $G = g_s(e^2/h)\sum_{n=1}^{N} T_n(\mu)$, with $g_s$ as the spin degeneracy of the electron (thus taking a value of 2) and $e$ as the elementary charge of the electron. The term $\sum_{n=1}^{N} T_n(\mu)$ of $G$ represents the electron transmission probability through $n$ individual quantum channels communicating with the QDs by an energy perturbation imposed by the electrode owing to the modulation of the electric field, stating for a function that models the nature and strength of the electron coupling of the QD states with the electrode.

\begin{figure}[t!]
\centering
\includegraphics[height=8.3cm]{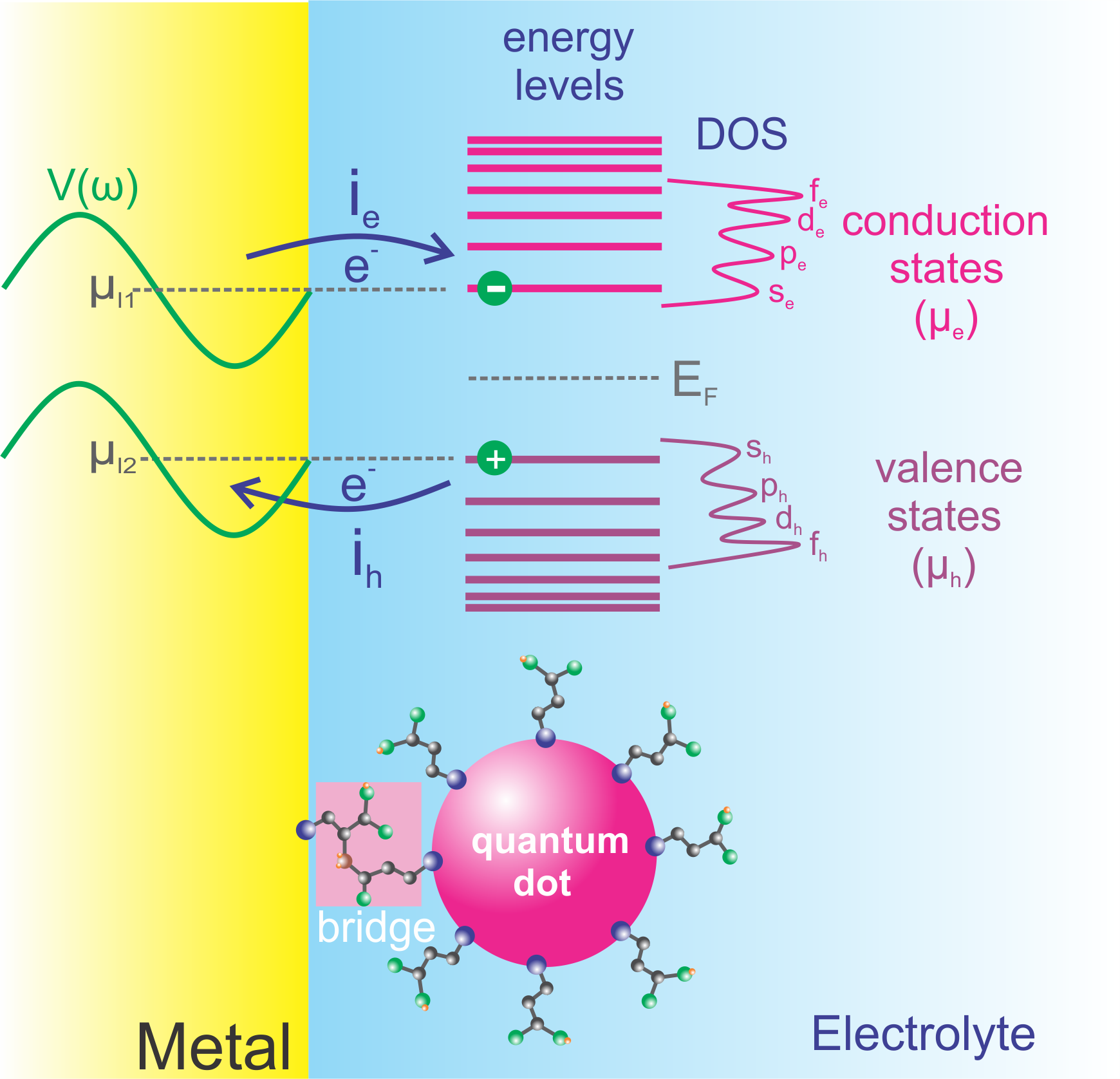}
\caption{Schematic representation of a semiconducting dot structure contacted to a gold electrode through an L-cysteine self-assembled monolayer. This interface is embedded into an electrolytic medium where the electron communication with a rate $\nu$ occurs from the electrode to the QD (upper picture) and \textit{vice-versa} (bottom picture) under a sinusoidal electric field perturbation from the electrode. This perturbation allows the measurement of the resonance electric current between the electrode states and the discrete energy states of the QDs. The resonant electric current is due to electron ($i_e$) and hole ($i_h$) charge carriers.}
\label{fig:electrode-QDs}
\end{figure}

Whenever the transmittance of the information from the electrode to the QD states and \textit{vice-and-versa} occurs in the ideal (adiabatic) mode, i.e., particularly when $\sum_{n=1}^{N} T_n(\mu)$ is unitary (meaning that the communication between the electrode and the QDs occurs adiabatically), the conductance $G$ is maximum and so it allows quantum mechanics to operates with particular settings such as $G$ equal to the conductance quantum $G_0 = g_s(e^2/h) \sim$ 77,5 $\mu$S\cite{Bueno2020-2}, meaning that the transport or the transmittance operates adiabatically. 

Particularly, $C_\mu$, in Eq.~\ref{eq:nu-2}, is defined as the series combination of two capacitive contribution as given by $1/C_\mu = 1/C_e + 1/C_q$ (see Eq.~\ref{eq:nu-2}), where $C_e$ is the electrostatic contribution associated with the spatial separation of charge due to Coulomb's interactions and thus depends on the geometric features, and $C_q$ is the quantum-mechanical contribution related to the energy related to the work required for the occupancy of the quantum states (see Fig.~\ref{fig:electrode-QDs})\cite{Bueno2020} defined as $1/C_q = 1/e^2 (1/(dN/d\mu)_l + 1/(dN/d\mu)_r) $. Note that in the definition of $C_q$, $(dN/d\mu)_l$ corresponds to the DOS of the metal electrode and $(dN/d\mu)_r$ to the DOS of the low-dimensional QD structures. Furthermore, the DOS of the metallic side of the junction is hundreds of times higher in magnitude than that of the QDs, i.e. $1/(dN/d\mu)_l$ and, therefore, $C_q$ measured in this interface is directly proportional to $e^2(dN/d\mu)_r$\cite{Bueno2014}, i.e. to the DOS of the QD structures assembled at the interface of the electrode is serving as the spectroscopic probe.

Finally, the presence of an electrolyte (Fig.~\ref{fig:electrode-QDs}) containing counter-ions contributing to electric-field screening is required for experimental measurements and permits the crucial approximation ($C_e \sim C_q$)~\cite{Pinzon2021} to be stated in Eq.~\ref{eq:nu-2}, enabling QRS to be a suitable methodology for measuring the DOS of QDs. Therefore, the approximation $C_e \sim C_q$ in Eq.~\ref{eq:nu-2} implies an energy degeneracy, as discussed in section~\ref{sec:QRS-fundamentals}. Therefore, analyzing Eq.~\ref{eq:nu-2} under the situation of modeling an adiabatic single electron transport within the degeneracy of the states consideration, it can be experimentally evidenced that the electrolyte environment~\citep{Pinzon-2022} plays a key role for the degeneracy condition, where it is noted a superposition of $C_e$ and $C_q$ capacitive states owing to an appropriate electric-field screening mechanism of the electrolyte~\citep{Bueno-2023-3}. These two superimposed capacitive states implying that the electron charge is shared between two capacitive states (classical and quantum) within the same electric potential, i.e. $e/C_e \sim e/C_q$, conducting locally (which can be either redox sites undergoing electron transfer reaction~\citep{Bueno-2023-3} or electronic states in organic semiconductor compounds) to $C_e \sim C_q$ in Eq.~\ref{eq:nu-2}. Hence, the equivalent $C_\mu$ capacitance, in this situation, is set as $1/C_\mu = 2/C_q$ in Eq.~\ref{eq:nu-2}, a state that is interpreted as an additional degeneracy of energy (besides that of the spin of the electron) owing to $e^2/C_\mu = 2e^2/C_q$, where 2 is accounted as the $g_e$ contribution in the formulation of rate $k = G_0/C_\mu$ that reduces to~\citep{Alarcon2021}

\begin{equation}
 \label{eq:G/C-degeneracy}
	\nu = g_s g_e \left( \frac{e^2}{hC_q} \right) = g_e \left( \frac{G_0}{C_q} \right) = g_s g_e \left( \frac{E}{h} \right). 
\end{equation}

Eq.~\ref{eq:G/C-degeneracy} was successfully verified by experiments and theoretically demonstrated to comply with diffusionless ET dynamics of redox reactions~\citep{Alarcon2021}, also with the electrodynamics and molecular electronics of organic semiconductor compounds~\citep{Pinzon-2023} and with the relativistic electrodynamics of graphene~\citep{Bueno-2022}. Note that origin of $g_e$ was introduced in the previous section and does not requires further additional analysis.

These energy and electric current degeneracies imply that the electric current is ambivalent, as is the case of redox reaction dynamics~\citep{Bueno-2023-3} and the quantum electrodynamics of organic semiconducting compounds and graphene~\citep{Bueno-2022}. The implication that the ambivalent electric current phenomenon is equivalently observed in organic semiconductor compounds (as will be demonstrated further) and graphene~\citep{Bueno-2022} is owing to the quantum rate theory being applicable despite the existence of redox reaction, but with the presence of ambivalent electric current. For instance, in graphene, there are negative and positive charge carrier concentrations within a two-dimensional structure that conforms with electron and holes representation~\citep{Lopes2021} that are, in ET reaction processes, associated with oxidation and reduction currents, which is well-known and referred to as faradaic current.

In other words, Eq.~\ref{eq:G/C-degeneracy} implies only the consideration of additional $g_e$ degeneracy besides $g_s$ in the heterogeneous electron transport rate mechanism stated by Eq.~\ref{eq:nu-2}. Note that for the specific case of molecular electronics, there is an adiabatic transport of a single electron (that conforms with ballistic electrodynamics) with the meaning that $G$, in Eq.~\ref{eq:nu-2}, is equivalent to $G_0 = g_s e^2/h$. This boundary condition implies simply considering a single adiabatic electron transport ideal situation in which $\sum_{n=1}^{N}T_{n}\left( \mu \right) = N\exp \left(-\beta L \right) = N\kappa \sim 1$, or a non-adiabatic situation in which $\nu = \kappa g_e G_0/C_q$, with $\kappa$ explicitly considered in the expression of $\nu$. Note that $\kappa$ in $\nu = \kappa g_e G_0/C_q$ implies the consideration of a non-adiabatic electron transport which is modeled as taking $\kappa$ lower than unity, but higher than null for electron transmittance to occur. Therefore, the only differences between the dynamics existing in electrochemistry reactions and that of electron resonant transport within push-pull semiconducting compounds are regarding the adiabatic (for push-pull dynamics) or non-adiabatic (for redox reactions dynamics) settings, but both electrodynamics, follows a relativistic character as is predicted by the quantum rate model depicted in Eq.~\ref{eq:G/C-degeneracy}.

In adiabatic or non-adiabatic situations, it has been demonstrated experimentally that the charge transfer resistance is equivalent to the resistance quantum $R_q = 1/G_0 = h/g_se^2 \sim$ 12.9 k$\Omega$, which experimentally validates the quantum rate theory~\citep{Sanchez-QR-Rct} in both situations of ET or ballistic transport of electrons. The demonstration of a quantum limit value of $R_q$ for the transport of electrons is achievable because $\kappa$ (for non-adiabatic processes) can be experimentally measured as well as the number of channel $N$ that is obtained from the measurement of $C_q$~\citep{Sanchez-QR-Rct}. Therefore, besides quantum rate theory~\citep{Bueno2020-2, Bueno-2023-3, Bueno-book-2018} encompasses other electron transport models and approaches such as that developed by Marcus~\citep{Marcus-1964, Marcus-1985, Marcus-1993}, the advantage of the quantum rate theory consists in its easy experimental test and applications owing to all the fundamental parameters of the theory are measurable.

Particularly, the fulfilment of the $C_e \sim C_q$ condition permits a direct correlation between $G$ and $C_q$ through $\nu$, where $\nu$ is related to the characteristic time-scale $\tau$, such as $\nu = 1/\tau$\cite{Bueno2020} (see Eq.~\ref{eq:Cq-complex}). This energy degeneracy state (imposed by the presence of an electrolyte) corresponds to a particular setting of Eq.~\ref{eq:nu} (see Eq.~\ref{eq:G/C-degeneracy}) and permits the study of the electronic structure of low-dimensional materials at room temperature without the need for ultra-high vacuum, hence not only constituting the requirement for the QRS methodology but also making this method advantageous. This fundamental basis of QRS can be established by noting that $\nu = E/h = e^2/hC_q$, where the only variable to be measured is $C_q$. Since $C_q$ can be obtained by electronic spectroscopy means, as described in detail in the experimental section through Eq.~\ref{eq:Cq-complex}, QRS is quite simple experimentally, although it does require a careful assembly and contact of the QDs to the probe electrode.

To resolve any questions related to the principles of a spectroscopic method based on the quantum-rate concept, defined in Eq.~\ref{eq:G/C-degeneracy}, let us demonstrate the correspondence between the duality particle $E = e^2/C_q$ (or $\textbf{p}$) and $\textbf{k}$ ($\omega$) wave of the electron whenever there is a time-dependent perturbation on $C_q$ that corresponds to a coherent response of $\textbf{k}$ and \textit{vice-versa}. Hence, $C_q = e^2 (dn/dE)$ can be rewritten as

\begin{equation}
 \label{eq:dE/dn}
	\left( \frac{dE}{dn} \right) = \frac{e^2}{C_q},
\end{equation}

\noindent where $e^2/C_q = \hbar \textbf{c}_* \cdot \textbf{k} = E$ is the energy associated with $C_q$, according to Eq.~\ref{eq:Planck-Einstein} and premises of the quantum-rate theory. Evidently, $dE = \hbar \textbf{c}_* \cdot \textbf{k} dn$ and a time perturbation in $E$ corresponds to a time perturbation of $n = q/e = L/\lambda$ (where $L$ is the length of the quantum channel and $\lambda$ the wave length associated to this length within $n$ as an integer quantum number) such that

\begin{equation}
 \label{eq:dE/dt}
  \left( \frac{dE}{dt} \right) = \hbar \textbf{c}_* \cdot \textbf{k} \left( \frac{dn}{dt} \right),
\end{equation}

\noindent where $E$ and $n$ are the only time-dependent variables; $dE/dt = E(t)$ and $dn/dt = n(t)$. Therefore, an energy perturbation $E(t)$ of a quantum RC system as a function of time corresponds to a coherent response of quantum RC states $n(t)$ as a function of time. These time-dependent and related functions can be stated as $E(t) = \bar{E} + \tilde{E}$ and $n(t) = \bar{n} + \tilde{n}$, respectively, where $\tilde{E} = E_0 \exp \left(j\omega t \right)$ and $\tilde{n} = n_0 \exp \left(j\omega t - \phi \right)$ are the oscillatory perturbation of $E$ and the corresponding oscillatory response of $n$ to a defined $E$ perturbation. 

$E_0$ and $n_0$ are the amplitudes of the energy perturbation of the quantum RC state response, respectively, and $\phi$ is the coherent phase difference between energy perturbation and quantum RC state response. $\bar{E}$ and $\bar{n}$ denote steady-state levels of the system where the oscillatory perturbation is performed.

The aforedescribed analysis shows that $E(t) = E_0 \exp \left( j\omega t \right)$ and $n(t) = n_0 \exp \left( j\omega t - \phi \right)$ allows defining a complex $E^{*}(\omega)$ energy state function such as

\begin{equation}
 \label{eq:dE/dn-complex}
	E^*(\omega) = \left[ \frac{dE(t)}{dn(t)} \right] = \left( \frac{E_0}{n_0} \right) \exp \left(j\phi \right),
\end{equation}

\noindent wherein the real component of $E^*(\omega)$ is identified as

\begin{equation}
 \label{eq:dE/dn-real}
	\Re \left[ E^*(\omega) \right] =  \left( \frac{E_0}{n_0} \right) \Re \left[ \exp \left(j\phi \right) \right] = \hbar \textbf{c}_* \cdot \textbf{k}.
\end{equation}

As discussed previously, if $\omega$, in Eq.~\ref{eq:dE/dn-real}, is considered $\omega_0$, there is a limit for the energy $E_0 = e^2/C_q^0$ that corresponds to a limit of Eq.~\ref{eq:Cq-complex} in which $e^2/n_0C_q^0 = \hbar \textbf{c}_* \cdot \textbf{k}$. This previous analysis implies a correspondence between $e^2/C_q^0$ and $\hbar \textbf{c}_* \cdot \textbf{k}$ that allows for the measurement of the electronic structure of QDs \textit{in-situ} by using an electric time-dependent method.

The mechanical statistics consideration of the QRS method will be considered in the next section.

\subsection{The Thermal Broadening Analysis of the QRS Method}

Since $C_q = e^2/E$ is a quantum variable obtained at room temperature~\cite{Bueno-David-2016, Bueno-David-2019}, as is the case with optical spectroscopic methods, it requires additional statistical mechanical inputs to appropriately treat the experimental signal of $C_q$. Hence, we note that in Eq.~\ref{eq:nu}, $C_q$ was defined at the zero absolute temperature limit \cite{Bueno2016} ($T \rightarrow$ 0 K) and, therefore, a more realistic setting requires a finite-temperature treatment of the $C_q$ measured at room temperature. This can be attained by considering the thermal broadening of the energy associated with $C_q$, i.e. $E = e^2/C_q$. 

The appropriate statistical mechanics treatment for this is achieved by considering the grand-canonical ensemble~\cite{Bueno-David-2019, Bueno-David-2016} or more directly by using the Fermi-Dirac distribution function $f(eV) = [1 + \exp(-\beta eV)]^{-1}$ to compute the dynamical occupancy of $C_q$ states coupled to the electrode at different potential energy levels of the electrode $-eV = \mu - E_F$, where $\beta = \alpha/k_B T$ and $k_B$ is the Boltzmann constant. $\alpha$ is the only \textit{ad-hoc} term that was introduced in this equation for modeling intrinsic inhomogeneities during the fitting of the experimental DOS peaks to this model (see Fig.~\ref{fig:DOS}). $E_F$ is the Fermi level of the QDs electronically coupled to the electrode and $\mu$ is the chemical potential of the electrons perturbed in the probe electrode.

Considering that $E_F$ is pinned, the changes in the potential level of the electrode are directly associated with the variations in $\mu$ owing to a $V$ applied with respect to $E_F$ (which is constant) such as that $d\mu = -edV$ is obeyed. Hence, the capacitance can be written as $C_q = dq/dV = e^2(dn/d\mu)$, where $dq = -edn$ is the charge variation due to $d\mu$ variations, where $n$ is the number of electrons (or states) communicating between the electronically-coupled QDs ensemble and the electrode. By noting that $n = Nf$, where $N$ is the total electron occupancy availability and $df/d\mu = f(1-f)$, a thermally broadened expression for $C_q$ is obtained as \cite{Pinzon2021}
 
\begin{equation}
 \label{eq:Cq-thermal}
    C_q = e^2 N \beta f(1-f),
\end{equation}

\noindent where $C_q$, in Eq.~\ref{eq:Cq-thermal}, is a function that predicts the thermal broadening of energy $E = e^2/C_q$ and consequently of the DOS\cite{Bueno2014}. This $C_q$ function can be experimentally measured by the EIS method as described in the experimental section, particularly within the analysis of Eq.~\ref{eq:Cq-complex}. Thus, it enables a methodology for interpreting the QR spectra peaks, as will be demonstrated in the next section for the measured spectra of the CdTe QDs structures.

Nonetheless, before continuing our analysis on the using of QRS for resolving the DOS of QDs structures, some practical concerns must be observed to ensure the accuracy and reliability of the results. The first is regarding the chemical or physical immobilization of the QDs over the probe electrode, which  must be performed in a way that assures an appropriate electronic coupling and contact between electrode and QD states, allowing electrical contact and the formation of an appropriate junction that enables electronic communication between the electrode and attached QDs.

Additionally, as QRS methodology is essentially based on electrochemistry, the chemical stability of the electrolyte must be a concern within the applied range of potential. Still regarding the properties of the electrolyte and the physical principles of QRS operation, as discussed in section~\ref{sec:QRS-fundamentals}, the ability to access the DOS of QD nanoscale assemblies through QRS is primarily attributed to the electrical field screening performed by the electrolyte over the charge of the quantum states. Hence, the concentration of the electrolyte used in the analysis is important for a suitable electric field screening. According to the previous works about the application of QRS methodology in graphene\cite{Lopes2021}, organic molecules\cite{Pinzon-2023} and inorganic semiconducting nanofilms\cite{Pinzon2021}, electrolyte concentrations typically ranges from 1 to 10 mM.

\subsection{Resolving the DOS of Quantum Dot Assemblies}

In this section, the usefulness of the QRS as a tool for revealing the electronic structure of low-dimensional assemblies (besides graphene~\citep{Bueno-2022,Lopes-2023}) is demonstrated by resolving the DOS function of a particular type of semiconducting QDs. Accordingly, CdTe QDs were assembled on a gold surface through a molecular bridge electronic coupling in order to follow the methodology described in the experimental section to study the electronic structure of a CdTe QDs ensemble using the QRS method (summarized in Fig.~\ref{fig:EIS-methodology}). $C_q$ was obtained at the equilibrium potential\cite{sun2002electrochemical,wang2021electrochemical}, corresponding to the open circuit potential (OCP) of the interfaces (0.14 V \textit{versus} Ag|AgCl). Hence, from QRS measurements, $C_q$ was directly extracted using the capacitive Nyquist plot for which the diameter of the semi-circle corresponded to the value of $C_q$ at equilibrium communication dynamics. In other words, the low-frequency limit of the capacitive spectra, corresponding to $(\omega \rightarrow 0)$, as given by Fig.~\ref{fig:EIS-methodology}\textit{b}, allowed to obtain the $C_q$ values of 3.28 and 1.55 $\mu$F cm$^{-2}$ for two different QDs assemblies with the average nano-particle sizes of 2.23 and 3.27 nm, respectively (see more details in SI.3)

\begin{figure}[t!]
\centering
\includegraphics[height=15cm]{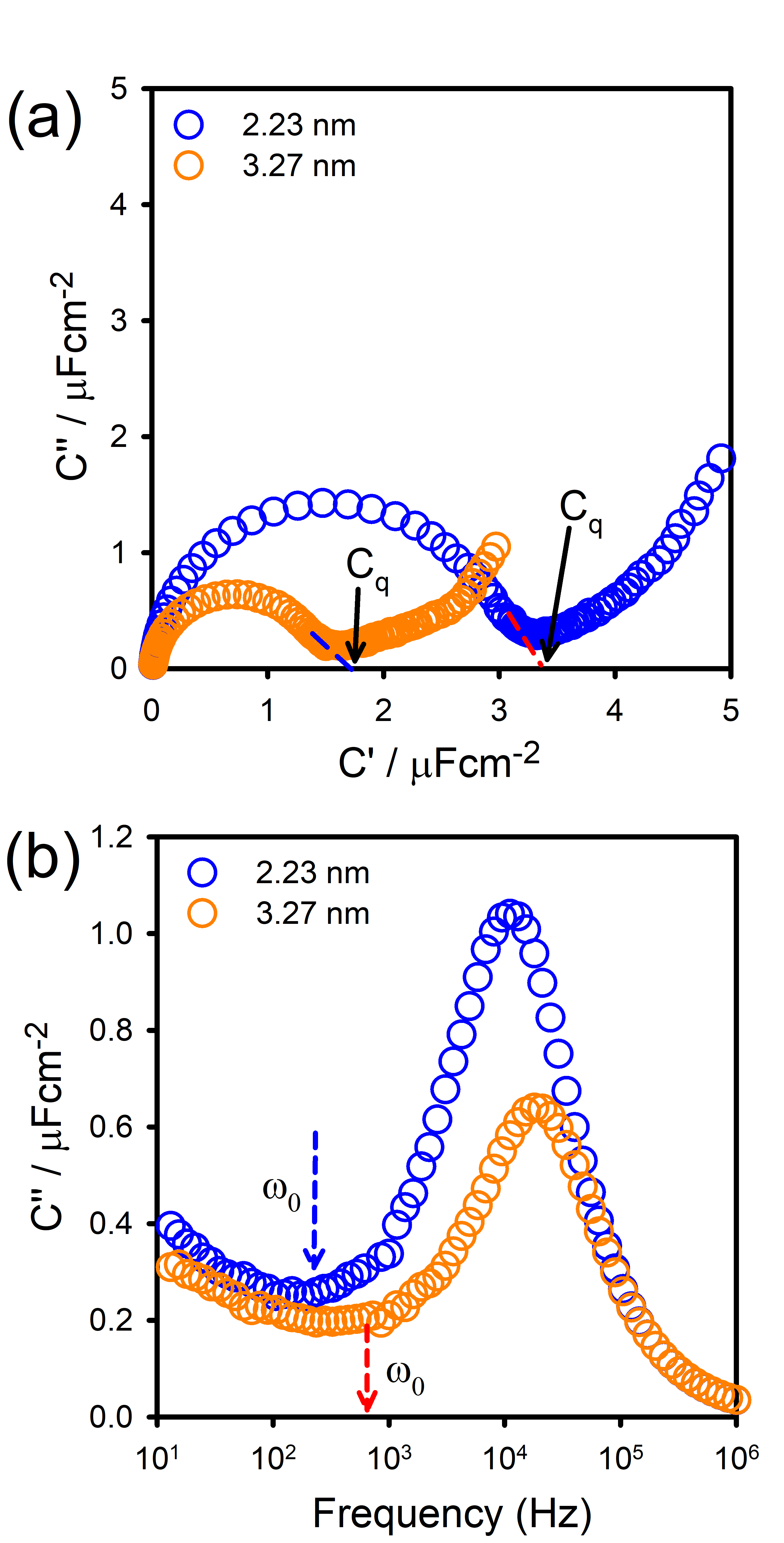}
\caption{(a) Capacitive Nyquist diagrams recorded for different QDs assemblies with two mean-average particle sizes (in a phosphate buffer solution with a pH of 7.4 at the equilibrium potential). The $C_q$ values for each of these assemblies were graphically obtained as the values of the semi-circle diameter, corresponding to two different equilibrium $\omega_0$ frequencies as indicated in the (b) Bode plots using the imaginary capacitance component of the spectra for these two QD assemblies.}
\label{fig:EIS}
\end{figure}

To measure the DOS, the equilibrium frequency $\omega_0$, as indicated in Fig.~\ref{fig:EIS}\textit{b}, was kept constant and a potential scan (from negative to positive) was applied to record $C_q$ as a function of the electrode potential. It is important to emphasize that the DOS measurements\footnote{Observe that there are some differences in the intensity of the peaks, but not in their potential positions.} are independent of scan perturbation direction, as depicted in Figure S2 of the SI document. This independence of scan direction and history of potential perturbation is owing to QRS consisting of an ultra-low energy perturbation methodology and consequently, this small potential perturbation allows electronic communication between the electrode and QDs assemblies without affecting the stability of the electronic structure.

The DOS function ($C_q \propto$ DOS) is obtained based on the $C_{e} \sim C_q$ approximation of the capacitive contribution, according to discussions conducted in a particular experimental setting of Eq.~\ref{eq:G/C-degeneracy}, as discussed in the section~\ref{sec:QRS-fundamentals}. The $C_q$ responses attained for the gold electrode and the L-cysteine monolayer in the absence of QDs are noteworthy. In the presence of QDs, as expected, the response is significantly distinct with regard to magnitude and shape, as can be noted in Figure S1 of the SI, demonstrating the successful immobilization of the QDs. The measurements, gold and L-cysteine,  attained in the absence of the QDs serves as a baseline reference of the signal.
    
\begin{figure}[t!]
\centering
\includegraphics[height=16cm]{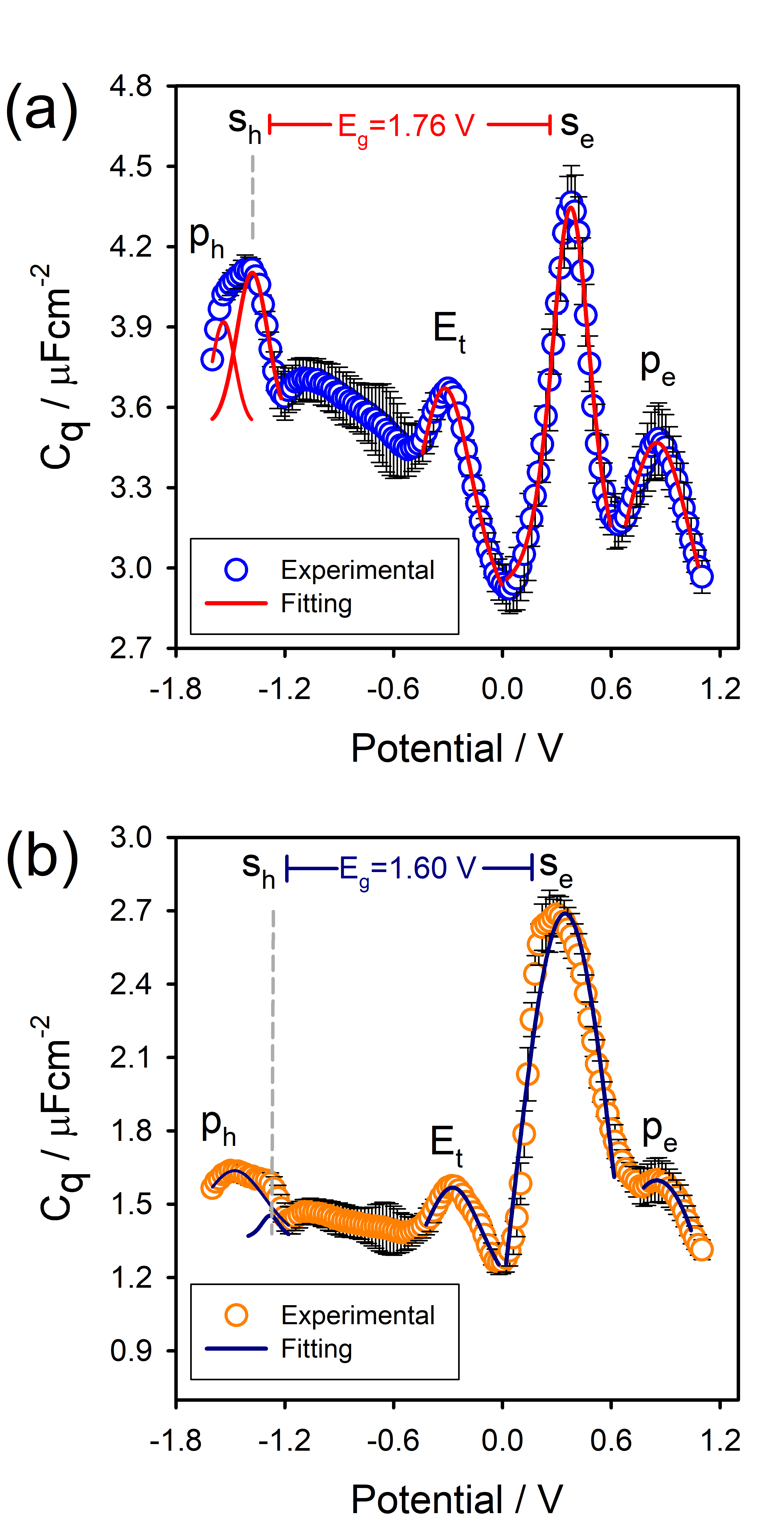}
\caption{Electronic DOS resolved from the measurements of the quantum capacitance $C_q$ for two different CdTe QDs assemblies with different nano-particle sizes, i.e. (a) 2.23 nm and (b) 3.27 nm. Each DOS plot is the average obtained for three independent QDs-interfaces and the error bar indicated the respective deviation. Potentials are reported \textit{versus} Ag|AgCl (3M KCl) reference electrode.}
\label{fig:DOS}
\end{figure}

Two sets of Gaussian-like functions separated by a roughly linear response were identified for both QD assemblies, as shown in Fig.~\ref{fig:DOS}\textit{a} and ~\ref{fig:DOS}\textit{b}. These responses were suitably fitted using the Eq.~\ref{eq:Cq-thermal} (red and dark blue line in Fig.~\ref{fig:DOS}), indicating good agreement between the experiments and the quantum rate theory (the fit R-squared coefficient of determination was $\sim$ 0.999). For an in-depth interpretation of the obtained DOS shape, it is important to recall the typical semiconducting QD electronic structures discussed in Fig.~\ref{fig:confinement}. This permits, in terms of quantum mechanical wave-function and atomic orbital representation, to note that the responses are correlated to symmetries that are referred to as s, p, d and so on~\cite{Vanmaekelbergh2005}. 

Therefore, valence and conduction band states of QDs are identified, respectively, as $s_h$, $p_h$ and $s_e$, $p_e$, where the subscript $h$ denotes hole states whereas $e$ refers to electron states. It is also important to highlight that for typical QD assemblies such as that depicted in Fig~\ref{fig:electrode-QDs}, individual QDs are weakly coupled to each other by van der Waals interactions owing to the capping molecules that separate the dots. Hence, the experimentally obtained Gaussian-like shape corresponds to the average response of the individual CdTe QDs in the ensemble\cite{Jdira2006} (see Fig.~\ref{fig:DOS}).

Accordingly, the Gaussian-like response corresponding to the most negative peak potential in Fig.~\ref{fig:DOS} and assigned at a peak potential of $\sim$ -1.4 V, is ascribed to the $s_h$ (HOMO) and $p_h$ valence states. These states were identified by using a deconvolution process as described in detail in the SI. Moreover, for both QD assemblies, it is observed that trapping states denoted as $E_t$ that are located in the band gap of the electronic structure  are present near the conduction band edge. These trapping states on average have maximum population at -0.30 V and -0.28 V, respectively, for the 2.23 nm (a) and 3.27 nm (b) nano-particle sizes, as depicted in Fig.~\ref{fig:DOS}. 

The presence of these trapping states was confirmed by independent photo-luminescence spectrum measurements that showed the presence of an additional emission besides the main peak expected in the photo-luminescence spectrum (as shown in SI.5). Finally, the peaks observed at more positive potentials in Fig.~\ref{fig:DOS} were ascribed to the conduction states $s_e$ (LUMO) and $p_e$. The electronic structure measured here using the QRS approach is in agreement with the DOS obtained for the CdSe and PbSe QDs assemblies via STS\cite{Liljeroth2005, Jdira2006}.

\begin{figure}[t!]
\centering
\includegraphics[height=15cm]{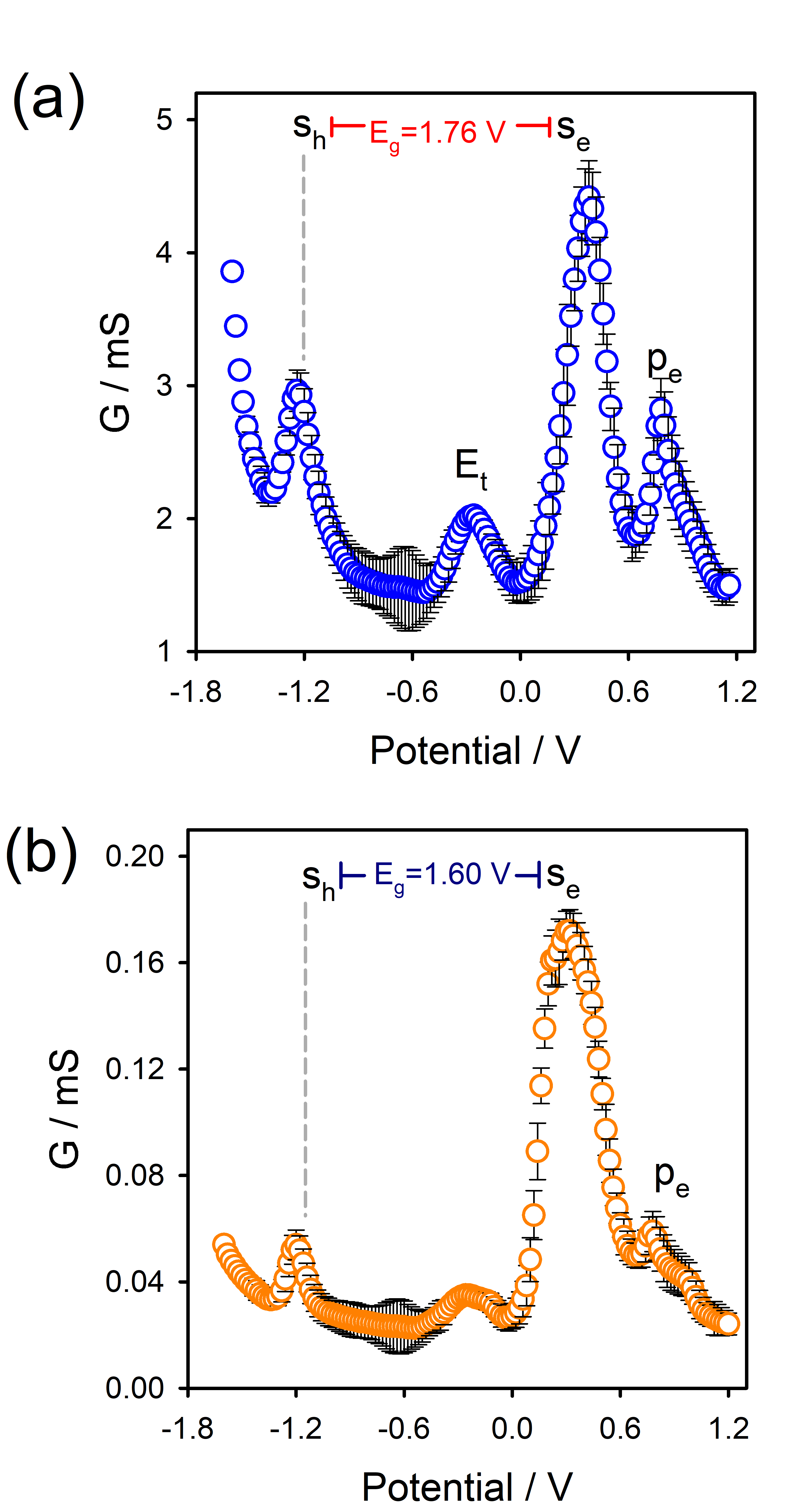}
\caption{Electronic DOS resolved from quantum conductance for two assemblies of CdTe QDs with different nanoparticle sizes of (a) 2.23 nm and (b) 3.27 nm. Each plot represents the average obtained from of the measurements over three independent electrodes and the error bar indicate their respective deviation. Potentials are reported \textit{versus} Ag|AgCl (3M KCl) reference electrode.}
\label{fig:conductance}
\end{figure}

For instance, in STS measurements, a potential difference is applied between a QDs-modified substrate and the microscope tip with the differences ascribed to the QDs electronic energy state levels within the electronic resonant (tunneling) currents established between the electrode-tip and QDs states with the transfer of electrons from the substrate (containing the assembly of QDs) to the tip or \textit{vice-versa}. The conductance response, as function of the potential energy, is obtained through these electric currents, where two sets of Gaussian-like shapes are identified in the STS measurements that are ascribed to the valence and conduction states of the QDs. The conductance response profile obtained by STS is comparable to the response of $C_q$ resolved by QRS (Fig.~\ref{fig:DOS}), demonstrating that both methodologies can access the electronic structure (DOS) of semiconducting QDs assemblies through a resonance electric response\footnote{Note that in the case of QRS, the resonance electric response obtained from a conductance plot (see also Fig.~\ref{fig:conductance}) is associated to a resonance electric current denoted previously here as $i_0 \propto C_q$ electric current, which is ambipolar in nature.} established between the electronic states and the probe (respectively tip and electrode in these two comparable methods of resolving the electronic structure of QDs). However, the DOS resolved by STS requires the use of expensive equipment as well as low temperature (5 K) and ultra-high vacuum conditions, whereas the DOS resolved by QRS is obtained at mild experimental conditions (room temperature and atmospheric pressure) due to the charge screening performed by the ions in the solvent-electrolyte environment, as introduced previously, in association with the meaning of $g_e$.

Additionally, based on the QR spectra shown in Fig.~\ref{fig:DOS}, the band gap $(E_g)$ energy can be estimated for these QD assemblies by calculating the differences between the $s_h$ (HOMO) and $s_e$ (LUMO) energy states. Hence, the band gap values of 1.76 and 1.60 $eV$ were obtained for the QD assemblies with the mean particle sizes of 2.23 nm and 3.27 nm, respectively. The higher band gap obtained for the smaller QD indicates a greater quantum confinement effect, in agreement with the expected correlation between the semiconducting nanocrystals band gap and the nanoparticle size\cite{Jdira2006}. Nevertheless, the band gap values estimated via the electrochemical QRS method are significantly lower than those obtained by optical approaches based on the absorption spectra of the same QDs in a solution environment (2.50 and 2.24 eV), as noted in Table SI.1 of the SI document. 

The discrepancy between these methods can be explained by considering that for a QDs solution, there are less interactions between the QDs and  the surrounding environment such as the electrode and capping molecules. Therefore, in the QRS method, the requirement for the attachment of QDs to the probe surface can be a source of errors in the quantitative estimation of the band gap energy. The electronic interaction between adjacent QDs in the assembly is noted when there exists a coupling of the electronic states between the adjacent QDs which is known as the quantum resonance phenomena between the dots~\cite{Lee2020, kim2015}. This phenomenon has been commonly observed in assemblies of QDs with short-chain capping molecules and short separation (less than 2 nm) between the neighboring QDs, such as the assembly of CdTe QDs capped by MPA-molecules (chain-length of 0.7 nm) studied here. Accordingly, the electronic coupling and communication between nearby QDs directly impacts the electronic properties of the assemblies, so that the values of the band gaps are smaller than for the condition of non-interacting QDs. These are the sources of errors or differences between the methodologies. It is clear that in the QR spectroscopic method, the preparation of the sample (assembly on the substrate) is important; however, this problem can be overcome by designing an experimental set-up in which the assembly will be molecularly coupled to the probe (electrode) in a suitably `diluted' way (with a reasonable distance established between adjacent QD moieties).

As shown in Figure \ref{fig:DOS}, each data point in the $C_q$ response represents the average value obtained from measurements performed on three independent QD interfaces. The error bars indicate deviations, which are quite small. Specifically, for the nanoparticle size of 2.23 nm, the magnitude for each QD size assembly varies from \num{1.92e-8} to \num{1.34e-7} $\mu$F cm$^{-2}$, and for the nanoparticle size of 3.27 nm, it varies from \num{0.81e-8} to \num{1.26e-7} $\mu$F cm$^{-2}$. These deviations are one or two orders of magnitude lower than the average value of the $C_q$ response. Therefore, this methodology allows for reproducible measurements between electrodes, demonstrating its potential as a faster and simpler tool for accessing the electronic structure of nanoscale assemblies in general.

Finally, it is important to note that according to Eq.~\ref{eq:G/C-degeneracy}, there exists a relationship between $G$ and $C_q$ through the rate $\nu$. This permits the construction of alternative methods for analyzing QRS spectra and validating the method. Therefore, due to this $\nu = G/C_q$ relationship, $G$ can be alternatively utilized to obtain information regarding the electronic structure of QDs. For instance, $G$ can be obtained as $G = \omega C''$ and a potential scan of $G$ at a fixed $\omega_0$ allows us to deduce the electronic structure and DOS pattern of the QDs similarly to using the $C_q$ signal of the response. Accordingly, following this reasoning, $G$ was obtained as a function of the potential, as depicted in Fig.~\ref{fig:conductance}. As expected, the $G-V$ pattern has a similar profile to that obtained for $C_q-V$ (Fig.~\ref{fig:DOS}). In summary, the obtained $G$ spectra confirmed the theory underlying the proposed QRS methodology and validated the method.

\section{Conclusions}
Electronic structures of assemblies of CdTe quantum dots were experimentally resolved from the measurements of the responses of the quantum capacitance $C_q$ and conductance $G$ using the quantum rate spectroscopic methodology in mild experimental conditions of ambient temperature and pressure in the presence of an electrolyte ambient. The methodology is based on the principles of quantum rate theory, as introduced in Eq.~\ref{eq:nu}, and is a novel tool that permits easier access to the information about the electronic structure of nanoscale materials, making it useful to the fields of physics, chemistry, nanoscience and nanotechnology. The results obtained for the specific case of CdTe quantum dots constitute an additional proof-of-concept of this spectroscopic methodology besides its use for measuring the electronic structure of graphene, as demonstrated previously~\cite{Lopes-2024}; hence, this method is not restricted to zero- or two-dimensional structures and can be also applied for the characterization different nanoscale structures of interest in different material science and chemistry fields.

\section*{Author Contributions}
\textbf{Edgar Fabian Pinzón Nieto}: Methodology, Investigation, Visualization. \textbf{Laís Cristine Lopes}: Methodology, Investigation, Visualization. \textbf{André Fonseca}: Methodology, Investigation, Visualization. \textbf{ Marco Antonio Schiavon}: Methodology, Investigation, Visualization. \textbf{Paulo Roberto Bueno}: Conceptualization, Supervision, Resources, Methodology, Writing - Original draft.

\section*{Conflicts of interest}
There are no conflicts to declare.

\section*{Acknowledgements}
This work was supported by the Sao Paulo State Research
Funding Agency (FAPESP) grants 2018/24545-9 and
2017/24839-0. The authors also acknowledge the support of the National Council for Scientific and Technological Development (CNPq) and Minas Gerais Research Funding Agency (FAPEMIG) for the grant PPM-00690-18.



\balance


\bibliography{rsc} 
\bibliographystyle{rsc} 

\end{document}